\pdfoutput=1 
\RequirePackage{ifpdf}	
\documentclass{JINST}

\title{Electrical Characterization of a Thin Edgeless N-on-p Planar Pixel
Sensors For ATLAS Upgrades}

\author{Marco Bomben$^1$\thanks{Corresponding author.},~
Alvise Bagolini$^2$, Maurizio Boscardin$^2$, Luciano
Bosisio$^3$, Giovanni Calderini$^{1,4}$, Jacques Chauveau$^1$, Gabriele
Giacomini$^2$, Alessandro La Rosa$^5$, Giovanni Marchiori$^1$ and Nicola Zorzi$^2$\\
\llap{$^1$}Laboratoire de Physique Nucleaire et de Hautes \'Energies (LPNHE), Paris, France\\
\llap{$^2$}Fondazione Bruno Kessler, Centro per i Materiali e i Microsistemi (FBK-CMM), Povo di Trento (TN),
Italy\\
\llap{$^3$}Universit\`a di Trieste, Dipartimento di Fisica and INFN, Trieste, Italy\\
\llap{$^4$}Dipartimento di Fisica E. Fermi, Universit\`a di Pisa, and INFN Sez. di Pisa, Pisa, Italy\\
\llap{$^5$}Section de Physique (DPNC), Universit\`e de Gen\`eve, Gen\`eve, Switzerland\\
E-mail: \email{marco.bomben@lpnhe.in2p3.fr}}

\abstract{In view of the LHC upgrade phases towards the High Luminosity LHC (HL-LHC), the ATLAS
experiment plans to upgrade the Inner Detector with an all-silicon system. Because of its
radiation hardness and cost effectiveness, the n-on-p silicon technology is a promising
candidate for a large area pixel detector. The paper reports on the joint development, by LPNHE and FBK of novel n-on-p edgeless
planar pixel sensors, making use of the active trench concept for the reduction of the dead
area at the periphery of the device. After discussing the sensor technology, and presenting some sensors' simulation results, a complete
overview of the electrical characterization of the produced devices will be given.}

\keywords{Fabrication technology; TCAD simulations; Planar silicon radiation detectors; Slim edge Sensors}

\begin{document}

\section{Introduction}\label{sec:intro}

The High Luminosity LHC (HL-LHC) will begin collisions around 2024 and will provide a leveled  instantaneous luminosity of $5\times10^{34} {\rm cm}^{-2}{\rm s}^{-1}$, 
with the aim of delivering an additional 2500~fb$^{-1}$ to ATLAS over ten years.
By then the ATLAS experiment will be equipped with a completely new Pixel Detector .  
The innermost layer of the new pixel detector will integrate a fluence of about $10^{16}\, {\rm 1\, MeV\, n_{eq}}/{\rm cm}^2$ for an integrated luminosity of 3000 fb$^{-1}$ 
($\sim$~10 years of operation).
These harsh conditions demand radiation-hard devices and a finely segmented detector to cope with the expected 
high occupancy.
The new pixel sensors will need to have  
  high geometrical acceptance: the future material budget restrictions and  
 tight mechanical constraints require the geometric inefficiency to be less than  2.5\%~\cite{IBL}.
One way to reduce or even eliminate the insensitive region along the device periphery is offered by
 the ``active edge'' technique~\cite{bib:Kenney}, in which a deep vertical trench is etched along the device periphery throughout the entire wafer thickness, 
thus performing a damage free cut (this requires using a support wafer, to prevent the individual chips from getting loose). 
The trench is  then heavily doped, extending the ohmic back-contact to the lateral sides of the device: the depletion region can then extend to the edge without causing 
a large current increase.
 This is the technology that was chosen bt FBK-Trento and LPNHE-Paris 
 for realizing n-on-p pixel sensors with reduced inactive zone whose features and measurement results are reported in this paper.

The paper is organized as follows: 
in Section~\ref{sec:prod} the active edge technology chosen for a first production of n-on-p sensors is presented. 
Studies performed with TCAD simulation tools  helped in defining the layout  and making a first estimation of the charge collection efficiency 
expected after irradiation; they are presented in Section~\ref{sec:tcad}.
In Section~\ref{sec:perf} the results from the electrical characterization of the sensors will be shown.
Eventually, in Section~\ref{sec:concl} some conclusions will be drawn and future plans will be outlined.

\section{The active edge production}\label{sec:prod}

The sensors are fabricated on 100~mm diameter, high resistivity, p-type, Float Zone (FZ), \textless100\textgreater\, oriented, 200~${\rm \mu}$m thick wafers. 
The active edge technology is used, which is a single sided process, featuring a doped trench, 
extending all the way through the wafer thickness, 
and completely surrounding the sensor. For mechanical reasons, a support wafer is  needed.
 Both homogeneous (``p-spray'') and patterned (``p-stop'') implants have been used to prevent pixels shorting due to  the positive fixed charge present in the oxide.
A boron implant for  the ohmic contact to the substrate (``bias tab'') is also added. 
More details on the adopted active edge technique and on the production in general can be found in~\cite{bib:nim2012}.
Nine FE-I4 readout chip~\cite{bib:fei4} compatible pixel sensors were put on each wafer; each sensor consists of an array of 336~$\times$~80 pixels, at a pitch  
of 50~${\rm \mu}$m~$\times$~250~${\rm \mu}$m, 
for an overall sensitive area of 16.8~mm~$\times$~20.0~mm. 
The nine FE-I4 sensors differ in the pixel-to-trench distance (100, 200, 300, and 400~${\rm \mu}$m) and in the number of the guard rings (0, 1, 2, 3, 5, and 10)  
surrounding the pixel area. The sensor with 3 GRs and a 200~$\mu$m pixel-to-trench distance features two different GR designs, and 
each of them is repeated twice. 
A list of the different FE-I4 sensor versions is reported in Table~\ref{tab:fei4_devices}.

\begin{table}[tbp]
\begin{center}
\begin{tabular}{ccc}
name & \# of GRs & pixel-to-trench distance (${\rm \mu m}$) \\
\hline
S1 &  0 & 100 \\
S2 & 2 & 100 \\ 
S3 & 1 &100\\
S4 & 3 & 200\\ 
S5 & 3 & 200\\
S6 & 3 & 200\\
S7 & 3 & 200\\ 
S8 & 5 & 300\\ 
S9 & 10 & 400 
\end{tabular}
\end{center}
\caption{\label{tab:fei4_devices}The list of the different FE-I4 compatible-sensor layouts. 
Two different designs are envisaged for the sensor with 3 GRs and 200~$\mu$m pixel-to-trench distance. See text for more details.}
\end{table}

\section{TCAD simulations}\label{sec:tcad}

In order to explore and compare the properties of the design variations considered, numerical simulations were performed with TCAD tools from SILVACO~\cite{Silvaco}.
 2D structures  have been simulated, varying parameters like the number of GRs and the pixel-to-trench distance. 
 Among all the observables that could be studied, the charge collection efficiency (CCE), for simulated un-irradiated and irradiated 
 sensors,  was the more interesting.
To describe the radiation damage, an effective model based on three deep levels in the forbidden gap  was used~\cite{bib:Pennicard}. 
Radiation-induced interface traps at the Si-SiO$_{\rm 2}$ interface are also included in the simulation, as described in~\cite{bib:InterfaceRD50}.
 
\begin{figure}[tbp]
\begin{center}
\includegraphics[width=0.69\textwidth]{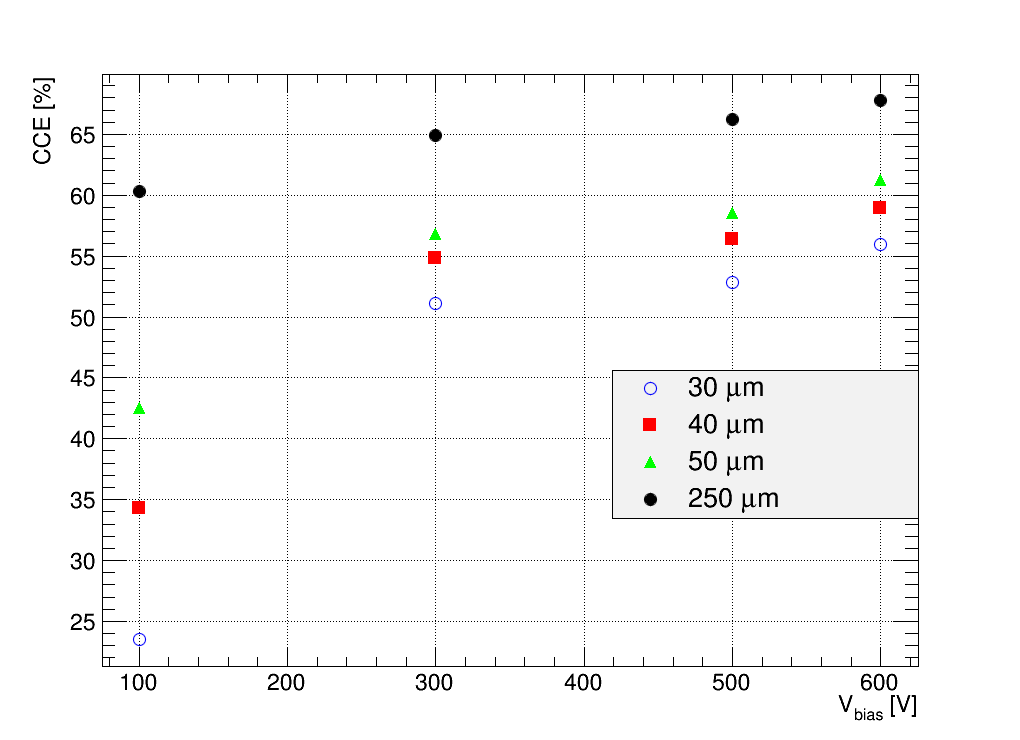}
\caption{\label{fig:CCE_comp_100_0gr}Charge collection efficiency as a function of bias voltage for an irradiated device at a fluence  $\phi= 10^{15} \rm{n_{eq}/cm^2}$ . 
The tracks are entering 
the detector either in the pixel region (``250~$\mu$m''  from the trench) or in the un-instrumented region (``30'', ``40'' and ``50~$\mu$m''  from the trench).
 The sensor has no GRs, and a 100 $\rm{\mu m}$ distance between edge and pixel.}
\end{center}
\end{figure}

To study charge collection efficiency (CCE) after irradiation,  charge creation in irradiated sensors was simulated. The most interesting case is when the charge is 
released in the gap between the pixel and the trench, when no GRs are present. If a significant amount of charge can be collected after irradiation in that
 region, the edgeless concept would be verified to work.
Our sensor was hit from the front side with  simulated tracks 
created by a minimum ionizing particle (MIP) traversing 200~$\mu$m of silicon ($\sim 2.6$~fC). 
The CCE was studied after a fluence of $\phi =  1 \times 10^{15} \rm{n_{eq}/cm^2}$ as a function of the bias voltage for the 
detector with no GRs and a 100~$\mu$m  trench-to-pixel distance. 
Four tracks impinging points  have been considered: three in the edge region, at 30, 40 and 50~$\mu$m  from the trench, and a fourth one in the pixel region, at a 
distance of 250~$\mu$m from the trench.
In the following the different incidence points will be identified by their entry point distance from the trench.
The charge collected by the pixel was defined as the integral over 100~ns\footnote{corresponding to 4 LHC bunch crossings} of the current flowing through the pixel, 
once the stable leakage current had been subtracted. Finally, the CCE was obtained by dividing this collected charge by the charge collected in the pixel region before 
irradiation~\footnote{This  normalization was chosen since it correspnds to non-edge region for a non-irradiated bulk}.
In Figure~\ref{fig:CCE_comp_100_0gr} the CCE is presented as a function of the bias voltage for the  simulated fluence for the four incidence points of the tracks. 

At a fluence  $\phi = 10^{15} \rm{n_{eq}/cm^2}$, for a bias voltage of 300~V more than 50~\% of the signal  is collected in the region that is 30~$\mu$m away from the trench; 
as a comparison, 65~\% of the  signal is retained in the  the pixel region.  The expected collected charge in the  region 30~$\mu$m away from the trench
is then of $\sim$8~ke~\footnote{the MPV for the charge created by a MIP in 200~$\rm{\mu m}$ is 16~ke}, 
which corresponds to  
a signal large enough to trigger the FE-I4 readout chip.

\section{Performance evaluation}\label{sec:perf}

\begin{figure}[tbp]
\begin{center}
\includegraphics[width=0.29\textwidth]{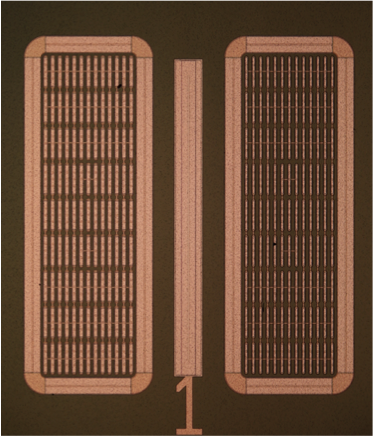}
\includegraphics[height=0.335\textwidth]{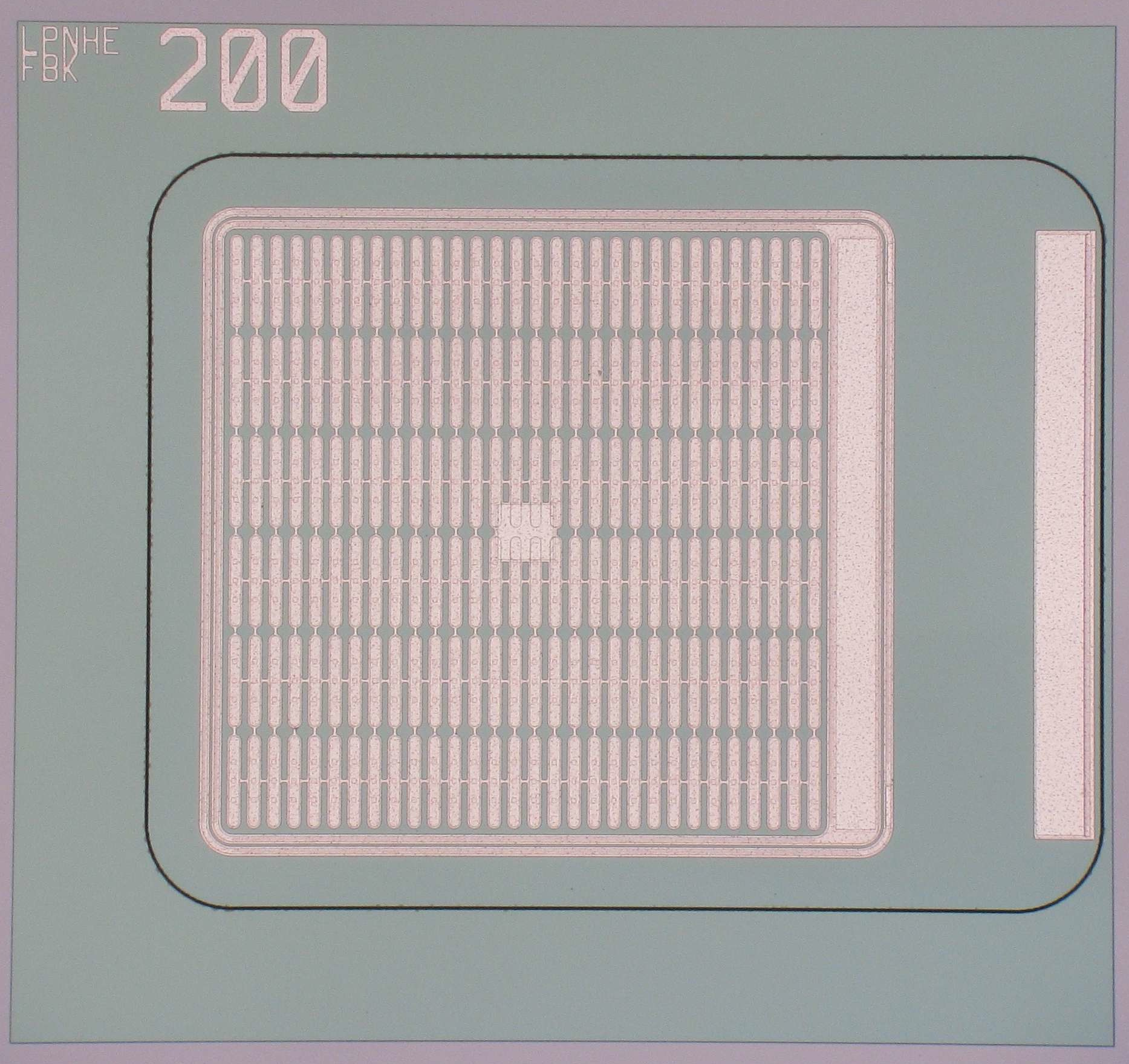}
\caption{\label{fig:pix_cap_struct}Left: test structures consisting of 2 arrays of 9~$\times$~13 FE-I4-like pixel cells each (``interpixel structure''); 
the pixels in the left (right) structure have (no) field-plate. 
Right: test structures consisting of an array of 6~$\times$~30 FE-I4-like
 pixel cells (``FE-I4 test structure''), where  all the pixels were shorted together were used to evaluate the current voltage characteristics  of the production.}
\end{center}
\end{figure}

Only one out of 20 processed wafers was not usable due to bad wafer-bonding.  
The electrical characterization of the production for non-irradiated sensors has been performed; it started with measurements on specially designed test structures, to assess
 mainly bulk and surface properties, then tests on large sensors followed.  
 The first part of the measurement program was carried mainly on structures reported in Figure~\ref{fig:pix_cap_struct}. 
 
 A test structure consisting of an array 
 of 9~$\times$~13 FE-I4-like pixel cells was used to measure the interpixel  and the pixel-to-backside capacitance; the central pixel was isolated with respect to all the other pixels;
 the first 8 neighbours were shorted together, but isolated from all the other remaining (which, again,  were shorted together). These structures are shown on the 
 left in Figure~\ref{fig:pix_cap_struct}, where two versions are present: one with metal field-plate and one without. ``Interpixel structure'' will be used for the sake of 
 brevity in the remaining of the text to refer to this structure.

In Figure~\ref{fig:pix_cap_struct}, on the right, an array of 6~$\times$~30 FE-I4-like
 pixel cells is shown; all the pixels were shorted together allowing the measurement of the current voltage characteristics  of the whole array and of the inner GR (if present), 
 and the BD voltage dependence on 
 the number of GRs and on the pixel-to-trench distance. Several combinations of values for the latter parameters are present on the wafer; in 
 Figure~\ref{fig:pix_cap_struct}, on the right, a structure with a 200~$\mu$m pixel-to-trench distance and 2 GRs is shown. ``FE-I4 test structure'' will be used for the sake of 
 brevity in the remaining of the text to refer to this structure.

\begin{figure}[tbp]
\begin{center}
\includegraphics[height=0.22\textheight]{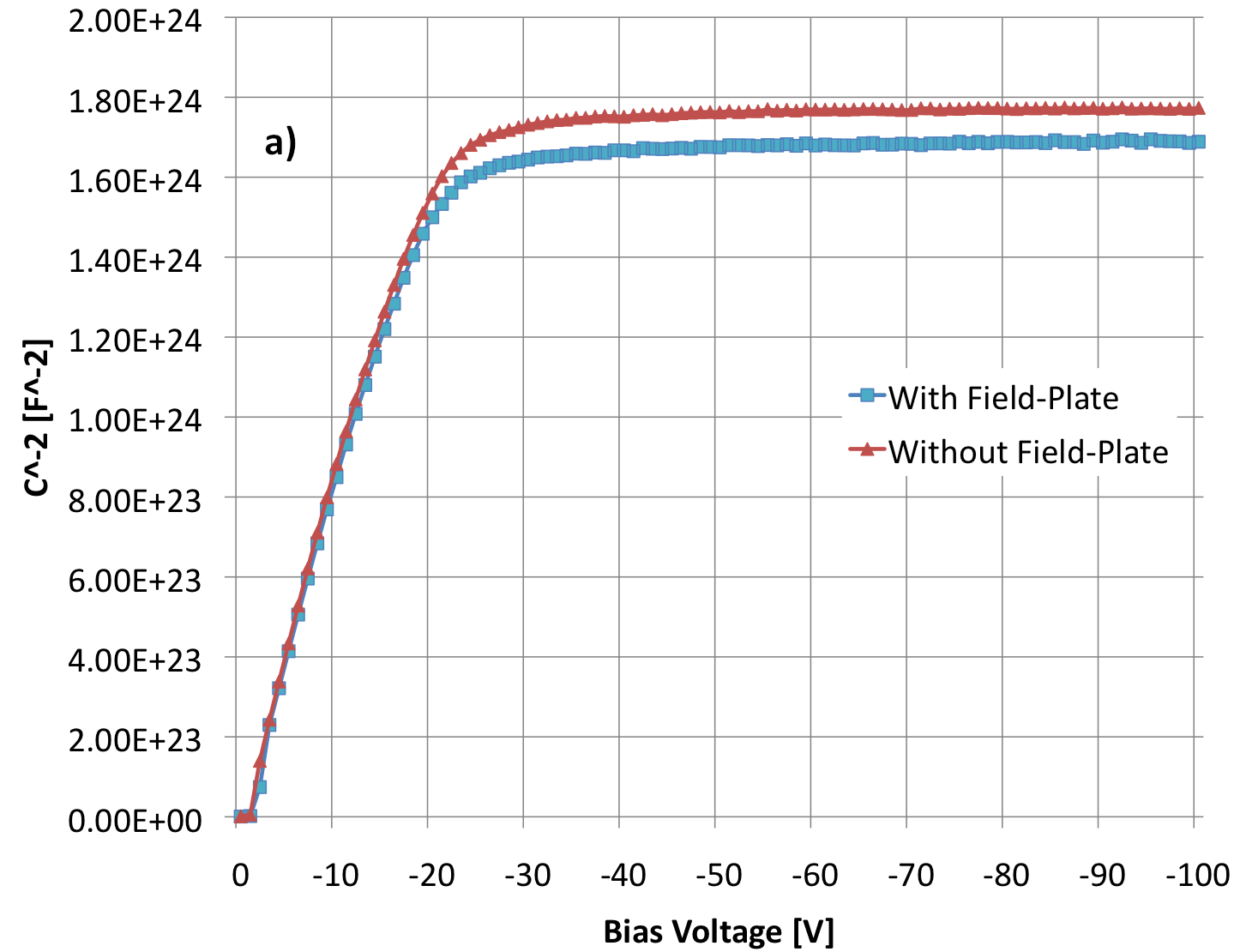}
\includegraphics[height=0.335\textwidth]{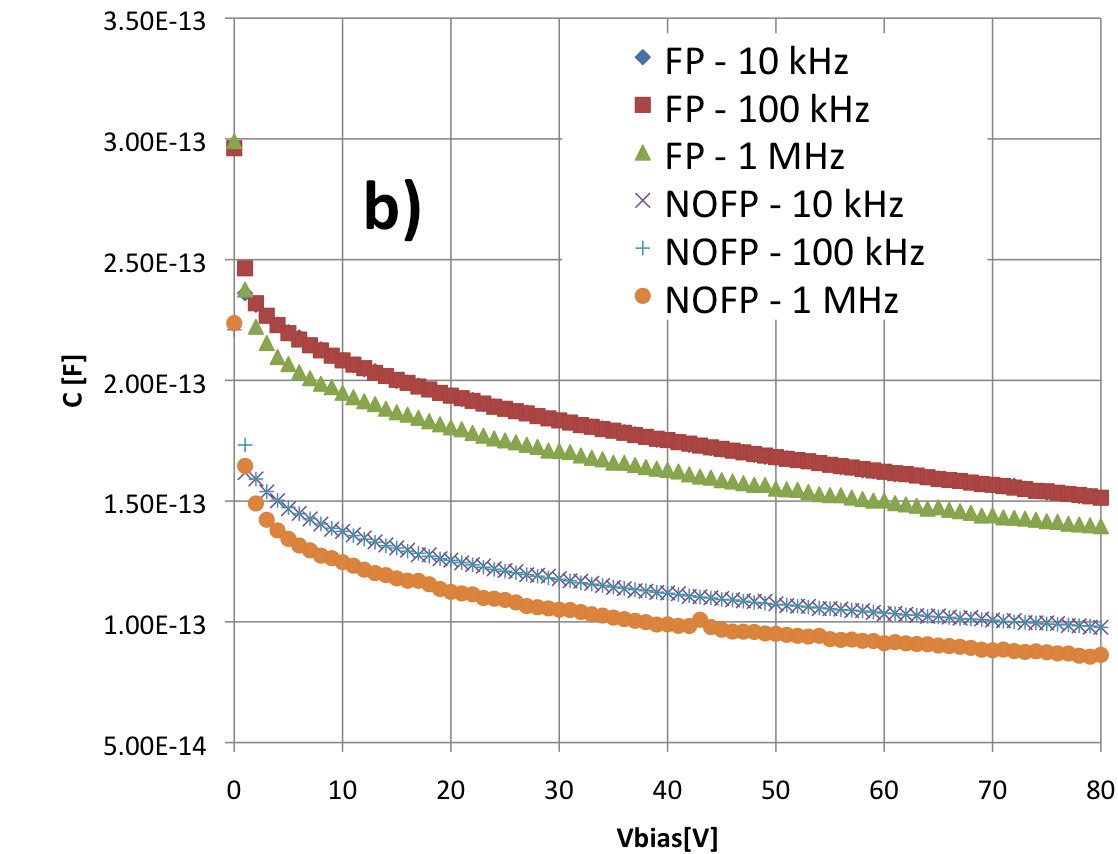}
\caption{\label{fig:cv-testpixels}Measurements results for the intepixel structure; a) inverse squared capacitance between all the  pixels and the sensor backside as a function of 
the bias voltage; both pixels with  and without field-plate were tested; b)
 interpixel capacitance for test structure with FEI4-like cells; the capacitance between the central pixel and
all the other pixels surrounding it in the test structure  is reported as a function of
the bias voltage for pixel cells with a field-plate, and
without it; the results are reported for three different frequencies: 10, 100~kHz and 1~MHz.}
\end{center}
\end{figure}

 For the interpixel structure, in Figure~\ref{fig:cv-testpixels} a) the inverse of the square 
 capacitance between all the pixels and the sensor backside is presented 
 as a function of the bias voltage; the measurement was performed at a frequency of 10~kHz. From this measurement the sensors' depletion voltage was derived ($\sim$20~V).
For the same structure, in Figure~\ref{fig:cv-testpixels} b) the capacitance between the central pixel and  all the other ones 
 is presented as a function of the bias voltage; the measurement has been carried out at three different frequencies: 10, 100~kHz and 1~MHz.
  It can be seen that the presence of a field-plate increases the interpixel capacitance. The coupling is particularly important due 
 to the presence of the uniform p-spray implant. However, the level of capacitative coupling, even with  a field-plate, is acceptable in term of electronic noise for the read-out.

\begin{figure}[tbp]
\begin{center}
\includegraphics[width=0.69\textwidth]{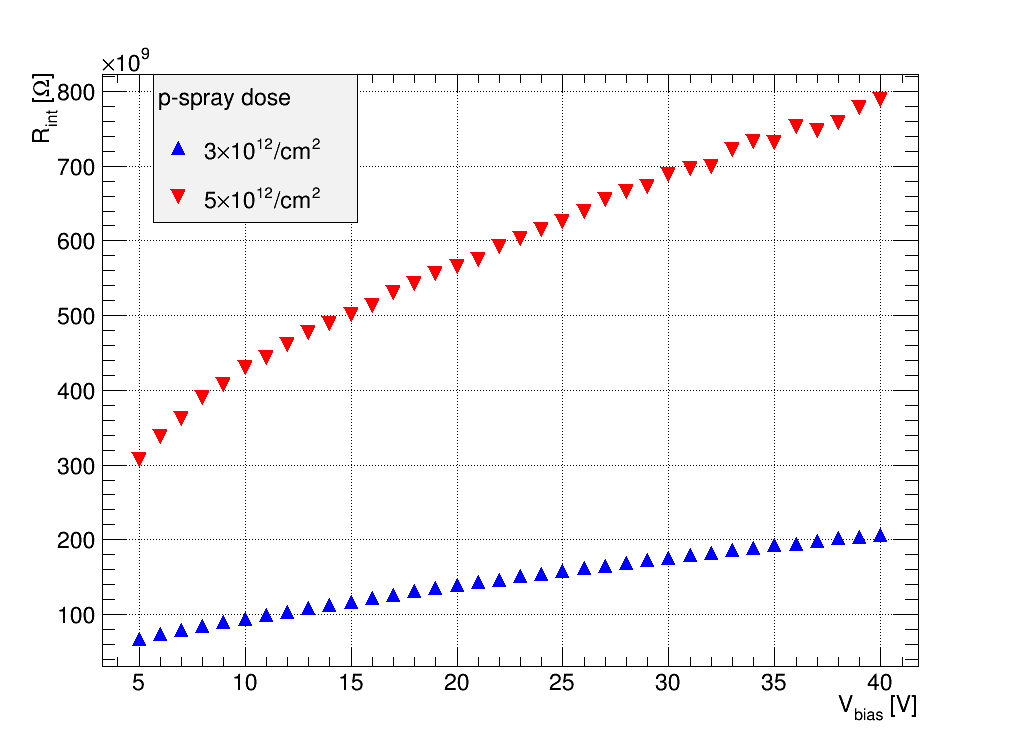}
\caption{\label{fig:Rint}Interpixel capacitance as a function of the bias voltage for two different p-spray doses.}
\end{center}
\end{figure}

Using the interpixel structure  the interpixel resistance was evaluated; the results are reported in Figure~\ref{fig:Rint} for the 2 different 
p-spray doses. It can be seen that for the high p-spray dose value, at depletion voltage, the interpixel resistance is four times larger than  the low 
p-spray dose corresponding value; nonetheless, excellent pixel isolation is already assured by low p-spray dose. 
After irradiation, this test will be  crucial  to prove the pixels isolation.

\begin{figure}[tbp]
\begin{center}
\includegraphics[width=0.75\textwidth]{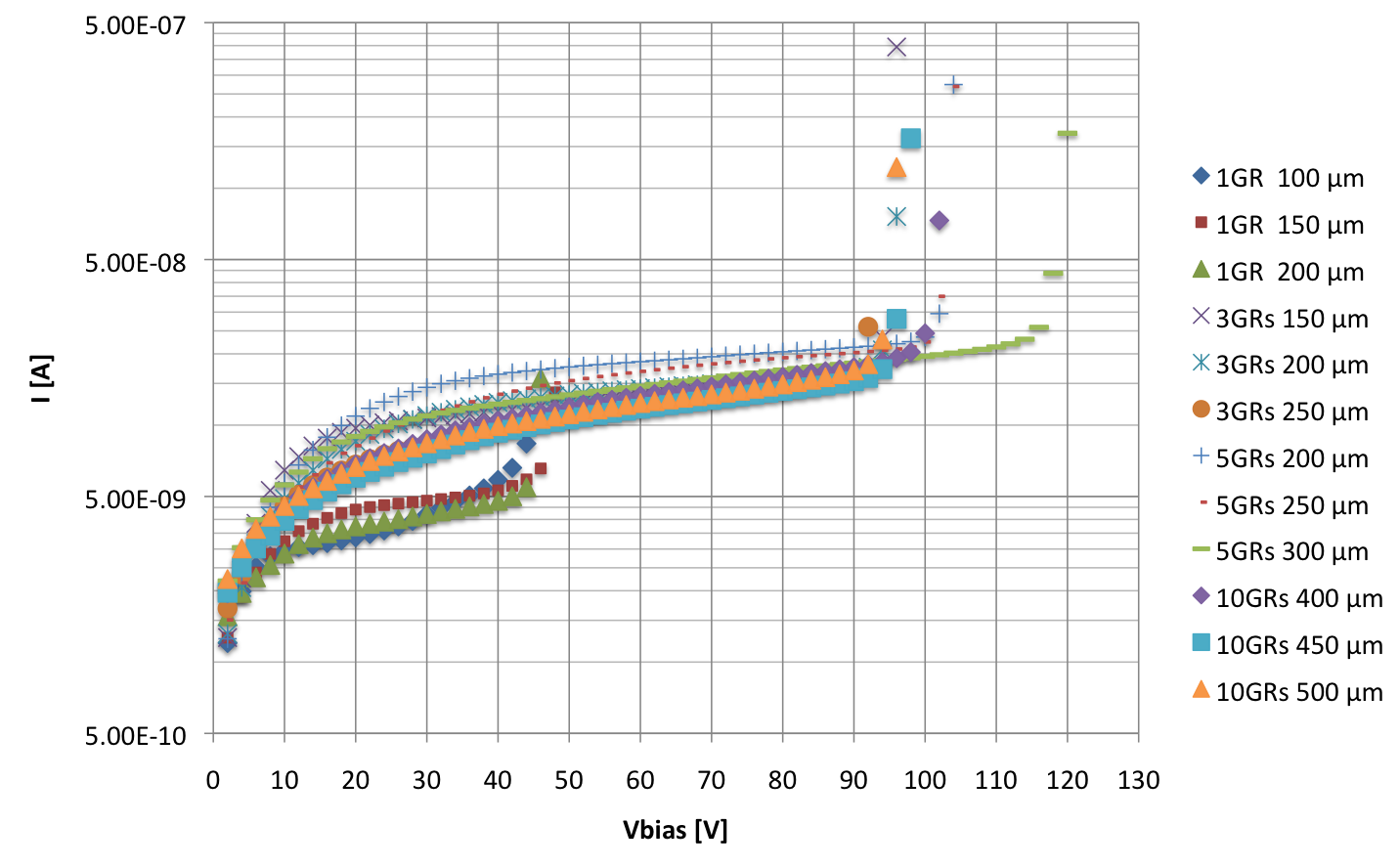}
\caption{\label{fig:iv-testpixels} Current voltage characteristics  for several FE-I4 test structures, 
differing by pixel-to-trench distance and by the number of GRs. The innermost GR was 
kept at ground (as well as the pixels) and  the current flowing through  is reported. Sensors were taken from a wafer featuring high p-spray dose.}
\end{center}
\end{figure}

 FE-I4 test structures  were used to evaluate the current voltage characteristics  of the production.
 The results are reported in 
 Figure~\ref{fig:iv-testpixels},  for 
the sensors from a wafer featuring a high p-spray dose; the devices were reversely biased via the bias 
tab,  the innermost 
GR was kept at ground (as well as the pixels), and the current flowing through the GR itself is reported. 

As it can be seen, adding more GRs increase the BD voltage and a wider edge-to-pixel distance corresponds to more bulk generated-current.  
All sensors can be operated in over-depletion (the measured depletion voltage was of $\sim$20~V).

\begin{figure}[tbp]
\begin{center}
\includegraphics[width=0.49\textwidth]{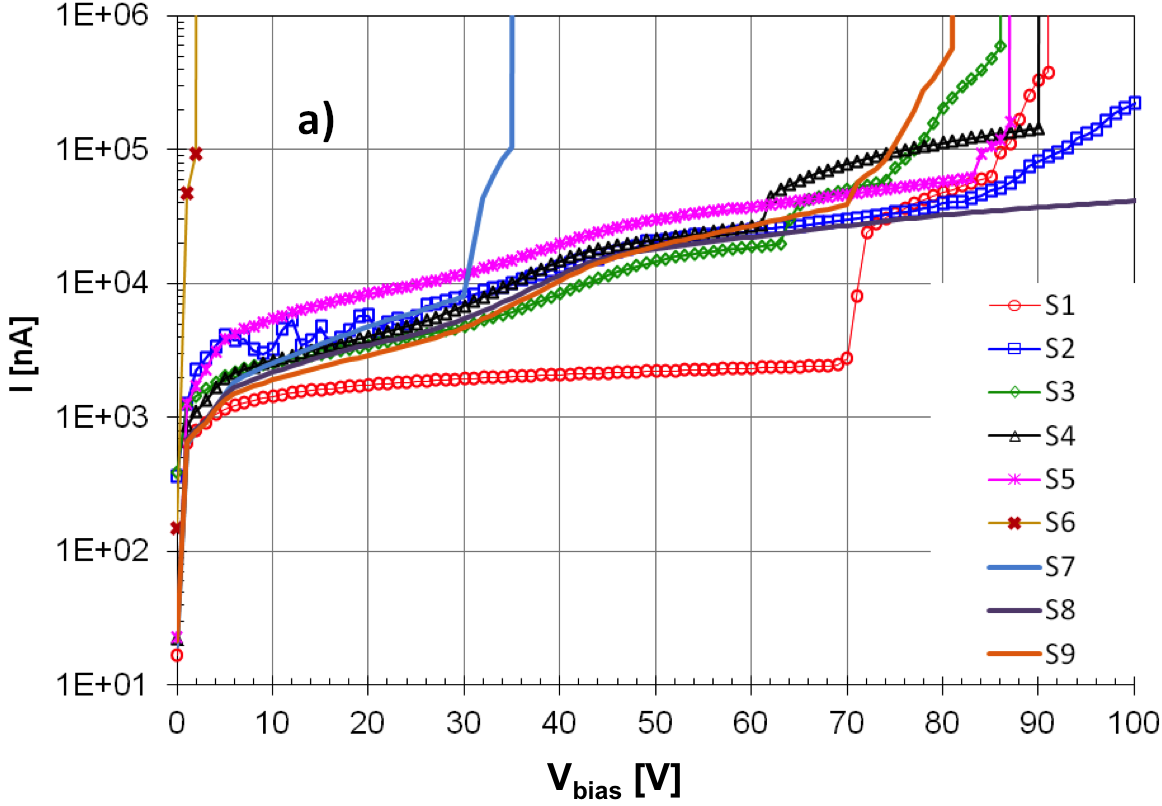}
\includegraphics[width=0.49\textwidth]{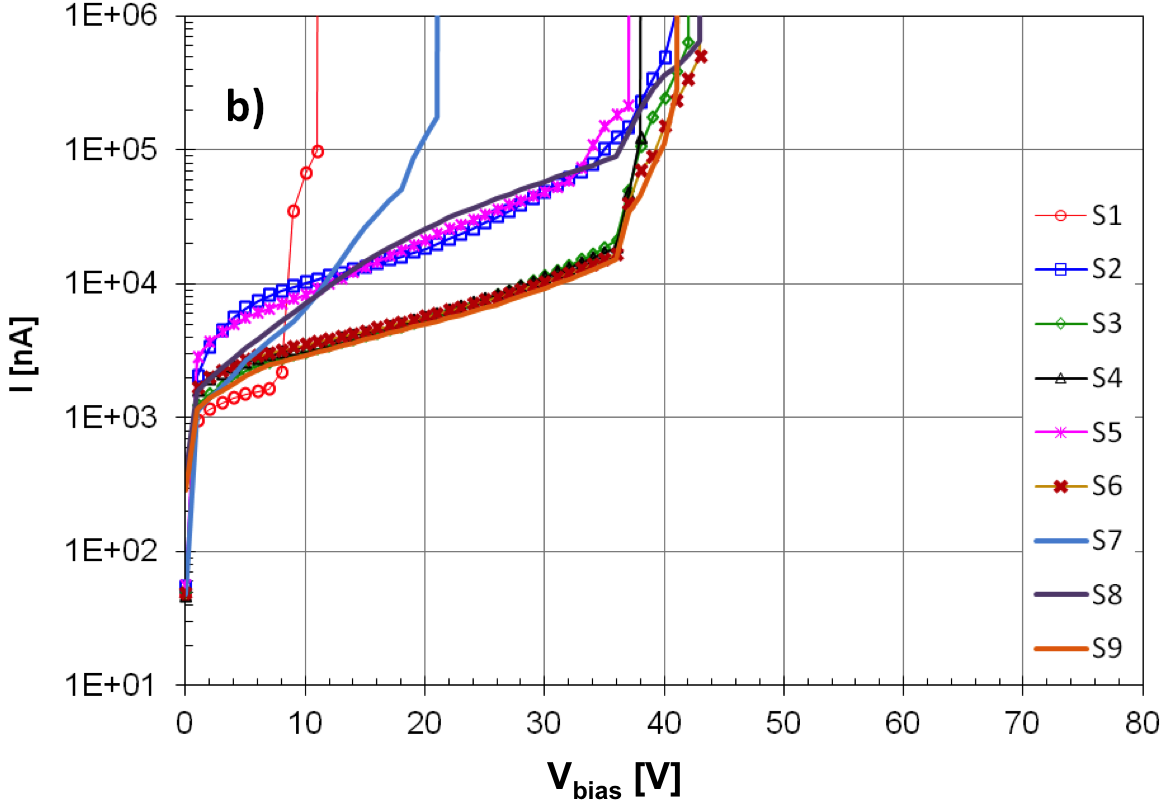}
\caption{\label{fig:FEI4}FE-I4 compatible sensors current voltage characteristics; a) p-spray dose on wafer was 3$\times$10$^{12}$/cm$^{2}$;  
right: b) p-spray dose on wafer was 5$\times$10$^{12}$/cm$^{2}$. Refer to text for the details on the different sensor layouts.}
\end{center}
\end{figure}

After having assessed the general production characteristics using the dedicated test structures,  the FE-I4 compatible pixels sensors current voltage characteristics  were 
measured. To this aim a temporary additional metal layer was used, as described in~\cite{bib:temporary}.
In Figure~\ref{fig:FEI4} the results are reported for two wafers, 
comparing sensors with two different p-spray implant doses; see Table~\ref{tab:fei4_devices} for the details on the different 
sensors' layouts. 
It can be seen that the higher the p-spray implant dose, the lower the BD voltage value; anyway, most of the sensors could be operated well in over-depletion. Large p-spray dose 
values are intended for after-irradiation operations, when larger oxide charges densities will make pixels isolation more challenging. These results are to be contrasted 
to the information contained in Figure~\ref{fig:Rint}: a large pixel dose assures an excellent pixel isolation, even after irradiation fluences comparable to those expected 
at the end of the HL-LHC phase, but they endanger operability due to low BD voltage. It must be said that anyway all but four defective sensors 
presented in Figure~\ref{fig:FEI4} can be 
operated in over-depletion and they are expected to show  larger BD voltages after irradiation.

\section{Conclusions and outlook}\label{sec:concl}

In view of the upgrade of the ATLAS Inner Detector for HL-LHC runs,
 FBK Trento and LPNHE Paris developed new planar n-on-p pixel sensors, characterized by a reduced inactive region at 
 the edge thanks to a vertical doped lateral surface at the device boundary, the ``active edge'' technology. 
 Simulation studies show the effectiveness of this technique in reducing the dead area, even after 
 simulated fluences comparable to those expected at the end of the HL-LHC phase for the external layers. 
The  measurements performed on real sensors, including capacitance- and current-voltage characteristics, show that it is possible to operate them successfully, well 
in over-depletion. Pixels isolation level is excellent even at low p-spray doses. Functional tests of the pixel sensors  with radioactive sources, before and after irradiation, 
 and eventually in a beam test, after having bump bonded a number of  pixel sensors to the FE-I4 read out chips, will follow. 

\acknowledgments

We acknowledge the support from the MEMS2 joint project of the Istituto Nazionale di Fisica Nucleare and Fondazione Bruno Kessler.


\begin{thebibliography}{9}


\bibitem{HL-LHC}
	L.~Rossi and O.~Br\"uning, \emph{High Luminosity Large Hadron Collider  A description for the European Strategy Preparatory Group}, CERN-ATS-2012-236, CERN, August 2012



\bibitem{IBL}
	ATLAS TDR 19, CERN/LHCC 2010-013, \emph{ATLAS Insertable B-Layer Technical Design Report},
	\href{http://cdsweb.cern.ch/record/1291633/files/ATLAS-TDR-019.pdf}{\tt http://cdsweb.cern.ch/record/1291633/files/ATLAS-TDR-019.pdf}




	


\bibitem{bib:Kenney}
	C.~J.~Kenney {\it et al.}, \emph{Results from 3-D silicon sensors with wall electrodes: near-cell-edge sensitivity measurements as a preview of active-edge sensors}, IEEE Trans. Nucl. Sci. {\bf NS-48 (6)} (2001) 2405.


\bibitem{bib:nim2012}
	M.~Bomben {\it et al.}, \emph{Development of Edgeless n-on-p Planar Pixel Sensors for future ATLAS Upgrades} Nucl.\ Instr.\ and Meth.  {\bf A 712} (2013) 41 - 47.

	


	
	

\bibitem{bib:fei4}
	M.~Garcia-Sciveres {\it et al.}, \emph{The FE-I4 pixel readout integrated circuit}, Nucl. Instr. and Meth. {\bf A  636} (2011) S155-S159.




\bibitem{Silvaco}
	\textit{Silvaco, Inc.}\\
4701 Patrick Henry Drive, Bldg 2\\
Santa Clara, CA 95054

\bibitem{bib:Pennicard}
	 D.~Pennicard {\it et al.}, 	\emph{Simulations of radiation-damaged 3D detectors for the Super-LHC}, Nucl. Instr. and Meth. {\bf A  592} (2008) 16-25.

\bibitem{bib:InterfaceRD50}
	J.~Schwandt {\it et al.}, \emph{Optimization of the radiation hardness of silicon pixel sensors for high x-ray doses using TCAD simulations}, 
	\jinst{7}{2012}{C01006}
	
\bibitem{bib:temporary}
	G.~Giacomini{\it et al.}, \emph{Development of Double-Sided Full-Passing-Column 3D Sensors at FBK}, IEEE Transactions on Nuclear Science 60,  (2013) 2357 

	

\end{thebibliography}
\end{document}